\documentclass[prl,twocolumn,showpacs,superscriptaddress,floatfix]{revtex4}
\usepackage{graphicx}

\newlength{\ldag}
\newcommand{\adag}{{a^\dagger}}
\newcommand{\ana}{{a^{\phantom\dagger}\hspace{-\ldag}}}

\newcommand{\rmd}{\mathrm{d}}
\newcommand{\pal}{\partial_l}

\def\nbN{\ensuremath{\mathrm{I\!N}}} 

\begin{document}

\title{Finite-Size Scaling Exponents of the Lipkin-Meshkov-Glick Model}
\author{S\'ebastien Dusuel}
\email{sdusuel@thp.uni-koeln.de}
\affiliation{Institut f\"ur Theoretische Physik, Universit\"at zu K\"oln,
Z\"ulpicher Str. 77, 50937 K\"oln, Germany}
\author{Julien Vidal}
\email{vidal@gps.jussieu.fr}
\affiliation{Groupe de Physique des Solides, CNRS UMR 7588, Campus Boucicaut,
140 Rue de Lourmel, 75015 Paris, France}

\begin{abstract}

We study the ground state properties of the critical Lipkin-Meshkov-Glick model.  Using the
Holstein-Primakoff boson representation, and the continuous unitary 
transformation
technique,  we compute explicitly the finite-size scaling exponents 
for the energy gap, the
ground state energy, the magnetization, and the spin-spin correlation 
functions. Finally, we
discuss the behavior of the two-spin entanglement in the vicinity of 
the phase transition.
\end{abstract}

\maketitle



Although Lipkin, Meshkov and Glick (LMG) introduced the model baring their
name in nuclear physics \cite{Lipkin65}, 
it is of much broader interest. It has thus been periodically revisited in different
fields, such as statistical physics of spin systems \cite{Botet82,Botet83} or
Bose-Einstein condensates \cite{Cirac98} to cite only a 
few. More recently it has also drawn much attention in the quantum information framework, 
where it has been shown to display interesting entanglement properties 
\cite{Vidal04_1,Vidal04_2,Vidal04_3,Latorre04_1} different from those observed in one-dimensional
models \cite{Osborne02,Osterloh02}.
After almost four decades, it has been proved to be integrable using
algebraic Bethe ansatz \cite{Pan99,Links03} or mapping it onto Richardson-Gaudin Hamiltonians for
which exact solutions have been proposed \cite{Dukelsky04}. However, although this
integrability provides some important insights about the structure of the spectrum, it is
useless to compute some physical quantities, such as the correlation functions, for a large
number of degrees of freedom. 
 
In the spin language that we shall adopt here, the LMG model describes
mutually interacting spins half, embedded in a magnetic field.
In the thermodynamical limit, it undergoes a quantum phase transition 
that is well described by a mean-field analysis. This transition can be first or 
second order depending whether the interaction is anti-ferromagnetic or ferromagnetic.
In the latter case and at finite size, some nontrivial scaling behavior of observables have been found
numerically \cite{Botet82,Botet83}. For instance, the energy gap seems to behave
as $N^{-1/3}$ at the critical point where $N$ is the number of spins.

In this letter, we explicitly compute these finite-size scaling exponents combining a $1/N$
expansion in the standard Holstein-Primakoff transformation, the continuous unitary 
transformations, and a scaling argument. 
First, we calculate the energy gap for which we detail the procedure. We
also give the leading finite $N$ corrections for the ground state 
energy, the magnetization, and the spin-spin correlation functions. 
In a second step, we discuss the two-spin entanglement properties through the concurrence
\cite{Wootters98} which is directly related to these functions. These latter results are in
excellent agreement with recent numerical studies \cite{Vidal04_1,Reslen04} which predict
a cusp-like behavior of the concurrence at the transition point.


We consider the following Hamiltonian introduced by LMG \cite{Lipkin65}
%
%
\begin{eqnarray}
H&=&-\frac{\lambda}{N}\sum_{i<j}
\left( \sigma_{x}^{i}\sigma_{x}^{j} +\gamma\sigma_{y}^{i}\sigma_{y}^{j} \right)
  -h \sum_{i}\sigma_{z}^{i} \\
&=&-\frac{2 \lambda}{N} \left(S_x^2 +\gamma S_y^2 \right) -2 h S_z +
{\lambda \over 2} (1+\gamma) \\
&=& - {\lambda \over N} (1+\gamma) \left({\bf S}^2-S_z^2 -N/2 \right)
-2h S_z \nonumber \\
&&  - {\lambda \over 2 N} (1-\gamma)\left(S_+^2+S_-^2\right)  \label{eq:ham},
\end{eqnarray}
%
%
where the $\sigma_{\alpha}$'s are the Pauli matrices,
$S_{\alpha}=\sum_{i} \sigma_{\alpha}^{i}/2$, and $S_\pm=S_x\pm i S_y$.
The $1/N$ prefactor ensures that the free energy per spin is finite in the thermodynamical
limit. Here we focus on the ferromagnetic case ($\lambda > 0$) and
all our results are valid for $|\gamma|\leq 1$ (the case $|\gamma| > 1$ 
being trivially obtained
by a simple rescaling of $\lambda$).
In this situation, the Hamiltonian (\ref{eq:ham}) displays a second-order
quantum phase transition at $\lambda=|h|$ \cite{Botet82,Botet83}.
In the sequel, we restrict our discussion to the  phase $|h| \geq 
\lambda$, and without loss
of generality, we set $h=1$.

The Hamiltonian $H$ preserves the magnitude of the total spin and does not
couple states having a different parity of the number of spins pointing in the
magnetic field direction (spin-flip symmetry), namely
%
\begin{equation}
\left[H,{\bf S}^2 \right]=0, \quad \mbox{and} \quad
\left[H,\prod_i \sigma_z^i \right]=0,
\label{eq:symmetries}
\end{equation}
%
for all values of the anisotropy parameter $\gamma$.
In the isotropic case  $\gamma=1$, one further has $[H,S_z]=0$, so that $H$ is
diagonal in the eigenbasis of ${\bf S}^2$ and $S_z$.
Due to the ferromagnetic interaction between the spins, the ground state
and the first excited state always lie in the subspace of maximum spin $S=N/2$.


In order to analyze the spectrum of $H$ in the large $N$ limit and to capture
the finite-size corrections, we
perform a $1/N$ expansion of the low-energy spectrum,  following the
ideas of Stein
\cite{Stein00}. We first use the Holstein-Primakoff boson
representation of the spin
operator \cite{Holstein40} in the $S=N/2$ subspace given by
%
   \begin{eqnarray}
     S_z&=&S-\adag a=N/2-\adag a, \label{eq:HP1}\\
     S_+&=&\left(2S-\adag a\right)^{1/2}a
     =N^{1/2}\left(1-\adag a/N\right)^{1/2}a, \label{eq:HP2}\\
     S_-&=&\adag\left(2S-\adag a\right)^{1/2}
     =N^{1/2}\adag\left(1-\adag a/N\right)^{1/2},
\label{eq:HP3}
   \end{eqnarray}
%
where the standard bosonic creation and annihilation operators satisfy $[a,\adag]=1$. This
representation is well adapted to the computation of the low-energy physics with $\langle\adag
a\rangle/N\ll 1$. After inserting these latter expressions of the spin operators in
Eq.~(\ref{eq:ham}), one expands the square roots in their Taylor 
series, and writes the result in 
normal ordered form with respect to the zero boson state. 
The Hamiltonian then reads $H=H_0+H_2^++H_2^-$, with 
%
%
\begin{equation}
   \label{eq:ham_HP}
   H_0=\sum_{\alpha,\delta\in \nbN}
   \frac{h_{0,\alpha}^{(\delta)} A_\alpha}{N^{\alpha+\delta-1}}
   \;\mbox{ and }\;
   H_2^+=\sum_{\alpha,\delta\in \nbN} \frac{h_{2,\alpha}^{(\delta)}
     \adag^2 A_\alpha}{N^{\alpha+\delta}},
\end{equation}
%
%
and with $H_2^-=\left({H_2^+}\right)^\dagger$ and 
$A_\alpha=\adag^\alpha\ana^\alpha$. 
The index $\alpha$ keeps track of the number of bosonic operators, 
and for a given $\alpha$, the superscript $\delta$ codes the successive 
$1/N$ corrections. 
For instance, the nonvanishing coefficients of $H_0$ are given by 
$h_{0,0}^{(0)}=-1$, 
$h_{0,0}^{(1)}=0$, $h_{0,1}^{(0)}=2-\lambda(1+\gamma)$,
$h_{0,1}^{(1)}=\lambda(1+\gamma)$ and $h_{0,2}^{(0)}=\lambda(1+\gamma)$.
%
%

Next, the Hamiltonian is diagonalized order by order in $1/N$ using the
continuous unitary transformation method, introduced by Wegner
\cite{Wegner94} and independently by
G{\l}azek and  Wilson \cite{Glazek93,Glazek94}. For a pedagogical
introduction to this technique, see
Ref. \cite{Dusuel04_2}. 
Note that the method has been applied to the LMG model in  
\cite{Mielke98,Pirner98,Scholtz03,Kriel04}, but its simultaneous use with 
the $1/N$ expansion originates in \cite{Stein00}. 
The main idea is to diagonalize the Hamiltonian in a continuous way starting from the
original (bare) Hamiltonian  $H=H(l=0)$.  A flowing Hamiltonian is then defined by
%
\begin{equation}
H(l)=U^\dagger(l) H U(l),
\label{eq:Hl}
\end{equation}
%
where $l$ is a scaling parameter such that $H(l=\infty)$ is diagonal.  A
derivation of Eq. (\ref{eq:Hl}) with respect to $l$ yields the flow equation
%
\begin{equation}
  \label{eq:dlH}
\pal H(l)=[\eta(l),H(l)], \mbox{ where } \eta(l)=-U^\dagger(l) \pal U(l).
\end{equation}
%
The anti-hermitian generator $\eta(l)$ must be chosen to bring the final
Hamiltonian to a diagonal form. Wegner proposed
$\eta(l)=[H_\rmd(l),H_\mathrm{od}(l)]=[H_\rmd(l),H(l)]$, where $H_\rmd$ and
$H_\mathrm{od}$ are the diagonal and off-diagonal parts of the Hamiltonian.
For our problem, it would read $\eta(l)=[H_0(l),H_2^+(l)+H_2^-(l)]$. Such a
choice suffers from the drawback that the tridiagonality of $H(l=0)$ is
lost during the flow and that $H(l)$ contains some terms which create any
even  number of excitations.
This problem can be circumvented using the so-called quasi-particle
conserving generator
$\eta(l)=H_2^+(l)-H_2^-(l)$ that we shall use here.  This generator
was first proposed in
\cite{Stein98,Mielke98} and given a  deeper physical meaning in
\cite{Knetter00}.

More generally, to compute the expectation value of any operator $\Omega$ on an
eigenstate $|\psi\rangle$ of $H$ with eigenvalue $E$, one must 
follow the flow of the
operator $\Omega(l)=U^\dagger(l) \Omega U(l)$, by solving
$\pal \Omega(l)=[\eta(l),\Omega(l)]$.
Indeed, one has:
%
\begin{equation}
\langle\psi|\Omega|\psi\rangle=\langle\psi|U(l=\infty)\: \Omega(l=\infty) \: 
U^\dagger(l=\infty)|\psi\rangle,
\end{equation}
%
where $U^\dagger(l=\infty)|\psi\rangle$ is  simply the eigenstate
of the diagonal Hamiltonian $H(l=\infty)$ with  eigenenergy $E$. 
In principle, one should follow the evolution of the $S_x$, $S_y$ and
$S_z$ observables, from which all others can be deduced. However, since we aim at computing
the ground state magnetization and spin-spin correlation functions, and because  of the
symmetries of the model, the calculation can be performed more simply as follows. 
First, the spin-flip symmetry (\ref{eq:symmetries}) implies 
%
\begin{eqnarray}
\langle S_x \rangle&=&\langle S_y \rangle=0,\\
\langle S_x S_z\rangle=\langle S_z S_x \rangle&=&\langle S_y
S_z\rangle=\langle S_z S_y \rangle=0.
\end{eqnarray}
%
Furthermore, since the maximum spin representation is
one-dimensional, the coefficients of the
eigenstates in this sector can be chosen to be real so that 
$\langle \{S_x,S_y \} \rangle=0$.
We are thus led to consider only the following (extensive) observables~:
$2S_z$, $4S_x^2/N$, $4S_y^2/N$ and $4S_z^2/N$. 
Of course, the structure of the 
flowing observables does not remain as simple as those of the initial conditions of the flow,
even with our choice of the generator.  In a notation similar to (\ref{eq:ham_HP}), all these
observables can be written as
$\Omega=\Omega_0+\sum_k(\Omega_k^++\Omega_k^-)$, where the sum runs over all
non-negative even integers $k$'s, and
%
\begin{equation}
   \label{eq:obs_HP}
   \Omega_0=\sum_{\alpha,\delta\in \nbN}
   \frac{\omega_{0,\alpha}^{(\delta)} A_\alpha}{N^{\alpha+\delta-1}}
   \mbox{ and }
   \Omega_k^+=\sum_{\alpha,\delta\in \nbN}
   \frac{\omega_{k,\alpha}^{(\delta)}
     \adag^k A_\alpha}{N^{\alpha+\delta+k/2-1}}.
\end{equation}
%
We have omitted the dependence on the flow parameter $l$ of the
$\omega$'s which is
implicit in the following. For instance, the initial conditions for $2S_z$ are
$\omega_{0,0}^{(0)}=1$, $\omega_{0,1}^{(0)}=-2$, with all other
coefficients vanishing.

The commutators $[\eta,H]$ and $[\eta,\Omega]$ are computed using
$[a,\adag]=1$ and basic counting results, yielding the flows
%
   \begin{eqnarray}
     \label{eq:flow_equations_h0}
     &&\pal h_{0,\alpha}^{(\delta)}=2\sum_{n,\alpha',\delta'}
     \mathcal{A}_{\alpha',\alpha-\alpha'-2+n}^{0,n} h_{2,\alpha'}^{(\delta')}
     h_{2,\alpha-\alpha'-2+n}^{(\delta-\delta'+1-n)},\quad\quad\\
     \label{eq:flow_equations_h2}
     &&\pal h_{2,\alpha}^{(\delta)}=\sum_{n,\alpha',\delta'}
     \mathcal{B}_{\alpha',\alpha-\alpha'+n}^{0,n} h_{2,\alpha'}^{(\delta')}
     h_{0,\alpha-\alpha'+n}^{(\delta-\delta'+1-n)},\\
     \label{eq:flow_equations_o0}
     &&\pal \omega_{0,\alpha}^{(\delta)}=2\sum_{n,\alpha',\delta'}
     \mathcal{A}_{\alpha',\alpha-\alpha'-2+n}^{0,n} h_{2,\alpha'}^{(\delta')}
     \omega_{2,\alpha-\alpha'-2+n}^{(\delta-\delta'+1-n)},\\
     \label{eq:flow_equations_ok}
     &&\pal \omega_{k,\alpha}^{(\delta)}=\sum_{n,\alpha',\delta'}
     h_{2,\alpha'}^{(\delta')}\big[
       \mathcal{A}_{\alpha',\alpha-\alpha'-2+n}^{k,n}
       \omega_{k+2,\alpha-\alpha'-2+n}^{(\delta-\delta'+1-n)}\nonumber\\
       &&\hspace{2.7cm}+\mathcal{B}_{\alpha',\alpha-\alpha'+n}^{k-2,n}
       \omega_{k-2,\alpha-\alpha'+n}^{(\delta-\delta'+1-n)}\big],
   \end{eqnarray}
%
with the definitions
%
\begin{eqnarray}
  \mathcal{A}_{\alpha',\alpha''}^{k,n}&=&n!\Big( C_{\alpha'}^n
    C_{\alpha''}^n-C_{\alpha'+2}^n
    C_{\alpha''+k+2}^n \Big),\\
  \mathcal{B}_{\alpha',\alpha''}^{k,n}&=&n! \Big( C_{\alpha'}^n
    C_{\alpha''+k}^n-C_{\alpha'+2}^n
    C_{\alpha''}^n \Big),
\end{eqnarray}
%
$C_\alpha^n$ being the binomial coefficient $\alpha!/[n!(\alpha-n)!]$.
The sums in (\ref{eq:flow_equations_h0}-\ref{eq:flow_equations_ok}) are
constrained by the fact that all subscripts and superscripts have to be
positive. For example, in (\ref{eq:flow_equations_h0}),
$n$ runs from $0$ to $1+\delta$,
$\alpha'$ from $0$ to $\alpha-2+n$
and $\delta'$ from $0$ to $\delta+1-n$.
At lowest nontrivial order in
$1/N$, Eqs. (\ref{eq:flow_equations_h0},\ref{eq:flow_equations_h2}) become
%
\begin{eqnarray}
  \label{eq:bogot_1}
  \pal h_{0,0}^{(1)}&=&-4 {h_{2,0}^{(0)}}^2\\
  \pal h_{0,1}^{(0)}&=&-8 {h_{2,0}^{(0)}}^2\\
  \label{eq:bogot_3}
  \pal h_{2,0}^{(0)}&=&-2h_{0,1}^{(0)} h_{2,0}^{(0)}.
\end{eqnarray}
%
These equations are the well-known Bogoliubov transform, written 
in a differential form \cite{Dusuel04_2}.

In \cite{Stein00}, Stein integrated analytically the flow equations
(\ref{eq:flow_equations_h0}) and (\ref{eq:flow_equations_h2}) for
$\gamma=-1$, and computed the $1/N$ corrections to the ground state energy
and to the gap. For our purpose, we have generalized his solution to any value of $\gamma$,
and went to order $1/N^3$. Furthermore, we also computed the exact solutions to
(\ref{eq:flow_equations_o0}) and (\ref{eq:flow_equations_ok}) and computed
the $1/N^2$ corrections to the extensive observables. Detailed calculations
will be presented elsewhere \cite{Dusuel04_3}.


To keep the presentation short, we will here only deal with the 
results obtained
for the energy gap $\Delta$. We found
%
\begin{eqnarray}
  \label{eq:gap}
  &&\frac{\Delta(N)}{\Delta(\infty)}=1+\frac{1}{N}\left[
    \frac{P_1(\lambda,\gamma)}{\Xi(\lambda,\gamma)^{3/2}}+\frac{Q_1(\lambda,
      \gamma)}{\Xi(\lambda,\gamma)}\right]\nonumber\\
  &&+\frac{1}{N^2}(1-\gamma)^2\left[
    \frac{P_2(\lambda,\gamma)}{\Xi(\lambda,\gamma)^{3}}+\frac{Q_2(\lambda,
      \gamma)}{\Xi(\lambda,\gamma)^{5/2}}\right]\\
  &&+\frac{1}{N^3}(1-\gamma)^2\left[ 
\frac{P_3(\lambda,\gamma)}{\Xi(\lambda,\gamma)^{9/2}}+\frac{Q_3(\lambda,\gamma)}{\Xi(\lambda,\gamma)^{4}}\right]+
\mathcal{O}\left(\frac{1}{N^4}\right),\nonumber
\end{eqnarray}
%
where $\Xi(\lambda,\gamma)=(1-\lambda)(1-\gamma\lambda)$,
$\Delta(\infty)=2 \: \Xi(\lambda,\gamma)^{1/2}$ is the mean-field
gap \cite{Botet82,Botet83}, and the $P_i$'s and $Q_i$'s are polynomial functions of
$\lambda$ and $\gamma$. For the 
isotropic case
$\gamma=1$, one has $P_1(\lambda,1)=4\lambda(1-\lambda)^2$,
$Q_1(\lambda,1)=-2\lambda(1-\lambda)$ and all contributions of order 
higher than $1/N$
vanish, so that we recover the exact result 
$\Delta_{\gamma=1}(N)=2(1-\lambda)+2\lambda/N$.

Let us now discuss the case $\gamma<1$ for which we have checked that 
$\lambda=1$ is
neither a root of the $P_i$'s nor of the $Q_i$'s. The result 
(\ref{eq:gap}) shows that all
$1/N^i$ corrections diverge when $\lambda$ approaches the critical 
value 1 in the
infinite system, such that the larger values of $i$, the  stronger the divergence.
However, physical  quantities cannot display any singularity
at finite $N$.  Using the usual ideas of finite-size scaling
\cite{Fisher72} generalized in
\cite{Botet82,Botet83} to infinitely coordinated systems, we can thus compute
the scaling critical exponents.
To this end, let us suppose $\lambda$ close to its critical value 1.
One can then neglect all
$Q$ terms which are less divergent than the $P$ ones.  In this limit,
expression (\ref{eq:gap})
becomes a function of the variable $N\Xi(\gamma,\lambda)^{3/2}$, namely
%
%
\begin{equation}
   \label{eq:scaling_function}
   \Delta(N)\simeq\Delta(\infty)
   \mathcal{F}_\Delta\left[N \: \Xi(\gamma,\lambda)^{3/2},\gamma\right],
   \mbox{ for } \lambda\simeq 1.
\end{equation}
%
%
Thus, the scaling function $\mathcal{F}_\Delta$ for the gap must behave as
$\left[N\Xi(\gamma,\lambda)^{3/2}\right]^{-1/3}$ in the vicinity of
the critical point $\lambda=1$,
for its  product with the mean-field gap to be non-singular.
Consequently, one gets
$\Delta(N)\sim N^{-1/3}$.
Of course we have only checked the scaling hypothesis up to the order
$1/N^3$, but the integrability
of the LMG model leads us to conjecture that the very simple
structure of the
$1/N$ expansion exhibited in (\ref{eq:gap}) is the same at all orders.

We have performed the same analysis for the ground state energy, the
magnetization and the two-spin correlation functions for the ground 
state. All results are summarized below and
detailed calculations will be presented in a forthcoming publication
\cite{Dusuel04_3}.
%
%
\begin{eqnarray}
\Delta(N) &\sim& a_\Delta N^{-1/3} \\
e_0(N) &\sim& -1-(1-\gamma)/(2N)+a_e N^{-4/3}\quad\quad\; \\
2\langle S_z   \rangle/N &\sim& 1+1/N+a_z N^{-2/3} \label{eq:sz}\\
4\left\langle S_x^2 \right\rangle/N^2 &\sim& a_{xx} N^{-2/3} \label{eq:sx2} \\
4\left\langle S_y^2 \right\rangle/N^2 &\sim& a_{yy} N^{-4/3} \label{eq:sy2}\\
4\left\langle S_z^2 \right\rangle/N^2 &\sim& 1+2/N+a_{zz} N^{-2/3} 
\label{eq:sz2}
\end{eqnarray}
%
%
where $e_0$ denotes the ground state energy per spin.
In each of the above expressions, we have first written the (exact)
non-singular contributions and, second, the term coming from the resummation
of the most singular terms in the $1/N$ expansion. Let us note however that
in (\ref{eq:sz}) and (\ref{eq:sz2}), the $N^{-2/3}$ terms dominate the large $N$ beha\-vior.
The coefficients $a$'s  are real numbers that cannot be  computed within our approach since
the scaling argument only provides the
exponents. Nevertheless, let us note that $a_{zz}=-a_{xx}$ since for
all $N$, one has
%
%
\begin{equation}
  \label{eq:S2}
  \frac{4}{N^2}\Big(\left\langle S_x^2 \right\rangle+\left\langle S_y^2
    \right\rangle+\left\langle S_z^2 \right\rangle\Big)
  =\frac{4{\bf S}^2}{N^2}
  =1+{2 \over N}.
\end{equation}
%
%
As $\left\langle S_x^2 \right\rangle$ is positive, one must also have
$a_{xx}\geq 0$.
One can furthermore infer that $N^{-4/3}$ corrections must exist in
$4\left\langle S_x^2 \right\rangle/N^2$ and/or in
$4\left\langle S_z^2 \right\rangle/N^2$, to cancel the one
of $4\left\langle S_y^2 \right\rangle/N^2$ in Eq.~(\ref{eq:S2}).

These results are in excellent agreement with the numerical data \cite{Botet83} where the
exponents for $\Delta$ and for $4 \left\langle S_x^2 
\right\rangle/N^2$ were conjectured.
However, the scaling exponent  2/3 for $2\langle S_z
\rangle/N$ differs from that found in \cite{Reslen04} $(0.55\pm 0.01)$. This discrepancy comes
from the too small system size investigated in \cite{Reslen04} ($N=500$ spins). 
We have indeed performed a numerical study up to $N=2^{14}$ spins and checked that the large $N$
leading exponent is indeed $2/3$ \cite{Dusuel04_3}.

The finite-size scaling of the correlation functions also allows us to discuss the entanglement
properties of the critical LMG model which have been  the subject of several
studies \cite{Vidal04_1,Vidal04_2,Vidal04_3}. For the ferromagnetic case  considered here, it has
been shown numerically that the two-spin entanglement, as measured by  the  (rescaled) concurrence
\cite{Wootters98}, displays a singularity at $\lambda=1$.
Actually, as shown
by Wang and M{\o}lmer \cite{Wang02}, the concurrence $C$ for
symmetric spin systems can be
simply expressed in terms of the spin-spin correlation functions.
More precisely, for the present case, one has 
%
%
\begin{equation}
  \label{eq:conc1}
(N-1) C= {2 \over N} \Big(\left|\left\langle S_x^2 -
S_y^2\right\rangle \right|-
N^2/4 +\left\langle S_z^2 \right\rangle \Big).
\end{equation}
%
%

At the critical point, using the results
(\ref{eq:sx2},\ref{eq:sy2}) and $a_{xx}\geq 0$, one can deduce that
$\left\langle S_x^2 - S_y^2\right\rangle$ is positive.
Then using (\ref{eq:S2}) and (\ref{eq:sy2}) one gets:
%
%
\begin{equation}
  \label{eq:conc2}
  (N-1) C_{\lambda=1}= 1-\frac{4\left\langle S_y^2\right\rangle}{N}
  \sim 1-a_{yy}N^{-1/3}.
\end{equation}
%
%
This behavior is in agreement with the numerical study of the
finite-size scaling presented in \cite{Vidal04_1,Reslen04}.
In the thermodynamical limit ($N \rightarrow \infty$), and in the 
phase $\lambda<1$, the Bogoliubov transform (\ref{eq:bogot_1}-\ref{eq:bogot_3}) also gives
\cite{Dusuel04_3}
%
%
\begin{equation}
  \lim_{N\to\infty}(N-1) 
C_{\lambda<1}=1-\sqrt{\frac{1-\lambda}{1-\gamma\lambda}},
\end{equation}
%
%
which generalizes to any anisotropy parameter the expression recently given by
Reslen {\it et~al.} \cite{Reslen04} for $\gamma=0$.


In summary, we have used the Holstein-Primakoff transformation and
the continuous unitary transformations to analyze the finite size 
corrections of
several observables in the LMG model.
Using a $1/N$ expansion and simple scaling arguments,
we have captured nontrivial exponents that had been conjectured since 
several decades (see
{\it e.~g.}  \cite{Botet82,Botet83}) but had never found any 
analytical support. This
powerful combination of both methods  clearly opens many routes to investigate.
In principle, the physics of the broken phase $(\lambda>1)$ could 
also be tackled using
the same approach, after performing a rotation bringing the $z$-axis 
along one of
the two directions of the classical magnetization. However, in this phase,
the gap, for instance, is known to behave like $\exp(-a N)$ 
\cite{Newman77} and may not be
extracted from a $1/N$ expansion. {\it A contrario}, the behavior of 
the other quantities
discussed here should be computed along the same line.

Finally, we wish to underline that the results presented here are
also relevant for the Dicke model \cite{Dicke54}. Indeed, in the zero temperature limit, the
LMG model can be put in a one-to-one correspondence with this latter model as
recently shown in \cite{Reslen04}.\\


\acknowledgments
We are indebted to B. Dou\c{c}ot, J. Dukelsky, S. Kirschner, D. Mouhanna, 
E. M\"uller-Hartmann, A. Reischl, A. Rosch  and K. P. Schmidt 
for fruitful and valuable discussions.
Financial support of the DFG in SP1073 is gratefully acknowledged.



\begin{thebibliography}{10}

\bibitem{Lipkin65}
H.~J. Lipkin, N. Meshkov, and A.~J. Glick, Nucl. Phys. {\bf 62},  188  (1965), 
N. Meshkov, A.~J. Glick, and H.~J. Lipkin, Nucl. Phys. {\bf 62},  199  (1965), 
A.~J. Glick, H.~J. Lipkin, and N. Meshkov, Nucl. Phys. {\bf 62},  211  (1965). 

\bibitem{Botet82}
R. Botet, R. Jullien, and P. Pfeuty, Phys. Rev. Lett. {\bf 49},  478  (1982).

\bibitem{Botet83}
R. Botet and R. Jullien, Phys. Rev. B {\bf 28},  3955  (1983).

\bibitem{Cirac98}
J.~I. Cirac, M. Lewenstein, K. M{\o}lmer, and P. Zoller, Phys. Rev. A {\bf 57},
   1208  (1998).

\bibitem{Vidal04_1}
J. Vidal, G. Palacios, and R. Mosseri, Phys. Rev. A {\bf 69},  022107  (2004).

\bibitem{Vidal04_2}
J. Vidal, R. Mosseri, and J. Dukelsky, Phys. Rev. A {\bf 69},  054101  (2004).

\bibitem{Vidal04_3}
J. Vidal, G. Palacios, and C. Aslangul, cond-mat/0406481.

\bibitem{Latorre04_1}
J.~I. Latorre, R. Or\'us, E. Rico, and J. Vidal, cond-mat/0409611.

\bibitem{Osborne02}
T.~J. Osborne and M.~A. Nielsen, Phys. Rev. A {\bf 66},  032110  (2002).

\bibitem{Osterloh02}
A. Osterloh, L. Amico, G. Falci, and R. Fazio, Nature (London) {\bf 416},  608
  (2002).

\bibitem{Pan99}
F. Pan and J.~P. Draayer, Phys. Lett. B {\bf 451},  1  (1999).

\bibitem{Links03}
J. Links, H.-Q. Zhou, R.~H. McKenzie, and M.~D. Gould, J. Phys. A {\bf 36},
  R63  (2003).

\bibitem{Dukelsky04}
J. Dukelsky, S. Pittel, and G. Sierra, Rev. Mod. Phys. {\bf 76},  643  (2004).

\bibitem{Wootters98}
W.~K. Wootters, Phys. Rev. Lett. {\bf 80},  2245  (1998).

\bibitem{Reslen04}
J. Reslen, L. Quiroga, and N.~F. Johnson, quant-phys/0406674.

\bibitem{Stein00}
J. Stein, J. Phys. G {\bf 26},  377  (2000).

\bibitem{Holstein40}
T. Holstein and H. Primakoff, Phys. Rev. {\bf 58},  1098  (1940).

\bibitem{Wegner94}
F. Wegner, Ann. Physik {\bf 3},  77  (1994).

\bibitem{Glazek93}
S.~D. G{\l}azek and K.~G. Wilson, Phys. Rev. D {\bf 48},  5863  (1993).

\bibitem{Glazek94}
S.~D. G{\l}azek and K.~G. Wilson, Phys. Rev. D {\bf 49},  4214  (1994).

\bibitem{Dusuel04_2}
S. Dusuel and G.~S. Uhrig, J. Phys. A {\bf 37}, 9275 (2004).

\bibitem{Mielke98}
A. Mielke, Eur. Phys. J. B {\bf 5},  605  (1998).

\bibitem{Pirner98}
H.~J. Pirner and B. Friman, Phys. Lett. B {\bf 434},  231  (1998).

\bibitem{Scholtz03}
F.~G. Scholtz, B.~H. Bartlett, and H.~B. Geyer, Phys. Rev. Lett. {\bf 91},
  080602  (2003).

\bibitem{Kriel04}
J.~N. Kriel, A.~Y. Morozov, and F.~G. Scholtz, cond-mat/0408420.

\bibitem{Stein98}
J. Stein, Eur. Phys. J. B {\bf 5},  193  (1998).

\bibitem{Knetter00}
C. Knetter and G.~S. Uhrig, Eur. Phys. J. B {\bf 13},  209  (2000).

\bibitem{Dusuel04_3}
S. Dusuel and J. Vidal, in preparation.

\bibitem{Fisher72}
M.~E. Fisher and M.~N. Barber, Phys. Rev. Lett. {\bf 28},  1516  (1972).

\bibitem{Wang02}
X. Wang and K. M{\o}lmer, Eur. Phys. J. D {\bf 18},  385  (2002).

\bibitem{Newman77}
C.~M. Newman and L.~S. Schulman, J. Math. Phys. {\bf 18},  23  (1977).

\bibitem{Dicke54}
R.~H. Dicke, Phys. Rev. {\bf 93},  99  (1954).

\end{thebibliography}

\end{document}